# Time Evolution of the Matter Content of the Expanding Universe in the Framework of Brans-Dicke Gravity

Sudipto Roy

Department of Physics, St. Xavier's College, Kolkata
30 Mother Teresa Sarani (Park Street), Kolkata – 700016, West Bengal, India.
email: roy.sudipto@sxccal.edu

**Abstract** A cosmological model has been constructed in the framework of Brans-Dicke (BD) gravity, based on an inter-conversion between matter and dark energy, for a spatially flat universe in the era of pressureless dust. To account for the non-conservartion of the matter content, a function of time f(t) has been arbitrarily put into the expression of the density of matter. By definition, f(t) is proportional to the matter content of the universe. Using suitable ansazes for scale factor and scalar field, the functional form of f(t) has been determined from the BD field equations. The scale factor has been so chosen that it would cause a signature flip of the deceleration parameter with time. The function f(t) decreases monotonically with time, indicating a transformation of matter into dark energy. Time dependence of the proportions of matter and dark energy of the universe has been determined. The effect of non-conservation of the matter content upon various cosmological parameters has been explored in the present study. Two models of matter-energy interaction have been proposed and f(t) has been expressed as a function of their interaction term. The dark energy equation-of-state (EoS) parameter has been expressed and analyzed in terms of f(t).

## 1 INTRODUCTION

On the basis of a large number of recent astrophysical observations it has been very firmly established that the universe is presently undergoing an accelerated expansion [Bennett et al. 2003; Riess et al. 1998; Perlmutter et al. 1999]. Several studies on cosmology have revealed that approximately seventy percent of the entire matter-energy content of the universe has a large negative pressure. This sector is known as dark energy, which is generally held responsible for the acceleration of the expansion process of the universe. Its true nature is yet to be ascertained. The general theory of relativity provides us with the cosmological constant ($\Lambda$), which is found to be one of the best candidates to represent dark energy and it is in agreement with observational findings sufficiently well, in spite of its own shortcomings [Sahni & Starobinsky 2000]. A number of models regarding dark energy have been formulated and their properties have been studied elaborately by the researchers [Sahni & Starobinsky 2000; Padmanabhan 2003]. Analyzing the data from supernova, a recent study has confirmed that there has been a signature flip of the deceleration parameter ($q$) of the universe, from positive to negative, indicating the existence of a phase of deceleration before the accelerated expansion began [Padmanabhan & Roy Choudhury 2003]. A phase of decelerated expansion of the universe preceded the present phase of accelerated expansion which is a very recent phenomenon [Roy Choudhury & Padmanabhan 2005]. It is important for the structure formation of the universe and its nucleosynthesis. According to the observational data and theoretical requirements, beyond a certain redshift (z) value the universe definitely had a phase of decelerated expansion [Riess et al. 2001; Goswami 2017]. It had a link with the generation of the dark energy content of the universe which has taken place in a manner such that its effect on the cosmic phenomena is more perceptible in the later phases of the matter dominated era.

In order to explain astrophysical observations regarding the accelerated expansion, many cosmological models have been proposed, without using the cosmological constant as the sole representative of dark energy [Copeland, Sami & Tsujikawa 2006; Martin 2008]. In all these models, the deceleration parameter has been found to be negative, showing the present phase of expansion to be an accelerating one. Models involving a scalar field with a positive potential are of great importance among these studies, where a negative effective pressure is generated if the potential term dominates the kinetic term. This scalar field is often called the quintessence field. A number of studies have been carried out regarding the quintessence potential and these results have been widely used. One may find a sufficient amount of information regarding this field in a study by V Sahni [Sahni 2004]. Many such potentials were worked out but they could not be properly explained by field theory [Banerjee, Das & Ganguly 2010].

It seems quite natural to think of an interaction between the main constituents of the universe, i.e. matter and dark energy, to account for the cosmological dynamics in a universe expanding with acceleration. This interaction may be assumed to play an important role in transferring energy from the field of one component to another. There are models according to which energy is transferred from the sector of dark matter to the sector of dark energy, leading to a predominance of dark energy over matter and causing the accelerated expansion during the later stage of cosmic evolution [Zimdahl 2012; Reddy & Kumar 2013].

To explain the transition of the universe properly, from a phase of decelerated expansion to its present phase of accelerated expansion, avoiding the discrepancies arising out of the arbitrariness in the formulation of above mentioned models, non-minimally coupled scalar field theories have been employed successfully. Here, a scalar field is incorporated in the theoretical framework, without bringing it separately in the theoretical discussions. The Brans-Dicke (BD) theory, which is considered to be the scalar-tensor generalization of the general theory of relativity, is the most spontaneous choice in this regard, due to its simplicity and its ability to generate results and predictions very close to those obtained from general relativity. Several features of the accelerating universe has been explained by BD theory and its modified versions [Banerjee & Pavon 2001; Brunier, Onemli & Woodard 2005]. The BD theory has been shown to generate sufficient acceleration in the matter dominated era even without considering the existence of any exotic quintessence field [Banerjee & Pavon 2001]. However, the scientific community has been searching for a theory to understand and analyze the change of cosmic expansion phase from deceleration to acceleration. There are models where dark matter and dark energy are treated as non-interacting entities having independent evolutions. One looks for a relatively generalized framework to analyze the behaviour of these two components of unknown nature, by considering an interaction between them. Two researchers, Zimdahl and Pavon, have shown that the cosmic coincidence problem can be effectively solved by using an interaction between dark energy and dark matter [Zimdahl & Pavon 2004]. The possibility of an inter-conversion of energy between the non-minimally coupled scalar field and the matter content of the universe had been predicted earlier by Amendola [Amendola 1999].

From several models, based on the Brans-Dicke (BD) theory, it has been obtained that the BD dimensionless parameter $\omega$ should have a very low value, typically of the order of unity, for an accelerated cosmic expansion [Sahoo & Singh 2002; Satish & Venkateswarlu 2014]. Assuming the scalar field to be interacting with the dark matter, it was once shown that a generalized form of the BD theory can account for an accelerated expansion even with a high value of the coupling parameter ($\omega$) [Banerjee & Das 2006]. In these studies, the BD theory is either modified to explain the findings properly or, one chooses a quintessence scalar field to account for the required acceleration. It has been shown by some studies that it is not necessary to use an additional potential to cause the deceleration parameter to change its sign from positive to negative, indicating a transition from deceleration to acceleration [Banerjee & Das 2006; Clifton & Barrow 2006]. To account for the observations connected to this phenomenon, they considered the dark matter and the BD scalar field to be interacting with each other. A recent cosmological model has explained the dynamics of the accelerating universe by considering an interaction between dark energy and dark matter [Das & Mamon 2014].

In this paper, a model has been proposed on the basis of generalized Brans-Dicke theory, where the coupling parameter $\omega$ is not a constant. It is a function of the scalar field ($\phi \equiv 1/G$) which varies with time. This generalized version of the BD theory was first proposed by Bergman and it was expressed in a more useful form by Nordtvedt [Bergman 1968; Nordtvedt 1970]. In the present study we propose a simple cosmological model on the basis of an assumption that the matter content of the universe is not conserved. Since dark energy is believed to be responsible for the accelerated expansion and, a phase of decelerated expansion preceded this accelerating phase, one may think of an increase of dark energy with time at the cost of the other main constituent of the universe, i.e. matter, in the present matter dominated era. The present model is based on BD field equations obtained for a homogeneous and isotropic (FRW) space-time, taking the pressure to be zero in a universe filled with pressureless dust. For this theoretical formulation, we have considered the possibilities of an inter-conversion between matter and dark energy. To account for the non-conservation of matter, we have incorporated a function of time $f(t)$ in the expression of the density of matter ($\rho$). The functional form of $f(t)$ $\left(\equiv \frac{\rho a^3}{\rho_0 a_0^3}\right)$ has been obtained from the BD field equations where $\rho$ stands for the density of matter in a universe with zero pressure [Banerjee & Ganguly 2009].

Section-2 of this study has the BD field equations obtained from a FRW space-time with zero spatial curvature for a dust filled universe with zero pressure. Section-3 of this article shows the derivation of the functional form of $f(t)$, using empirical expressions of the scalar field parameter ($\phi$) and the scale factor ($a$) in BD field equations. The function $f(t)$ has been found to be decreasing with time. Expressions representing the time dependence of the relative proportions of matter and dark energy of the universe have been obtained from $f(t)$, considering the matter content to be decaying into dark energy. We have also determined the time dependence of BD coupling parameter ($\omega$), the density of matter ($\rho$), and the gravitational constant ($G$). From these discussions one can determine the effect of decaying matter content upon the time variations of different cosmological parameters. In section-4 we have proposed two theoretical models showing the derivation of $f(t)$ as a function of the interaction term ($Q$) between matter and dark energy. The equation-of-state (EoS) parameter for dark energy has been derived as a function of $f(t)$. Section-5 contains graphical depictions and interpretations of all theoretical findings obtained in the previous sections. The time variations of several quantities of cosmological interest have been found to depend upon a parameter (denoted by $k$), which determines how fast the scalar field ($\phi$) changes with time. The findings of the present study regarding the time evolutions of some cosmological parameters are found to be in agreement with the results of some other studies of recent times, demonstrating the theoretical validity of this model. The present formulation provides us with a simple way to determine the time variations of matter and dark energy of the universe and also their relations with the time evolution of other parameters in the framework of BD theory.

## 2 BRANS-DICKE FIELD EQUATIONS

In the Brans-Dicke theory of gravitation the action is given by [Brans & Dicke 1961],

$$S=\int\left[\frac{\phi R}{16\pi G}+\frac{\omega(\phi)}{\phi}\phi_{,\mu}\phi^{,\mu}+L_m\right]\sqrt{-g}\;d^4x, \tag{1}$$

Here, $R$ denotes the Ricci scalar, $L_m$ stands for the matter Lagrangian, $\phi$ is the Brans-Dicke scalar field and $\omega$ is a dimensionless coupling parameter. In generalized Brans-Dicke theory $\omega$ is considered to be a function of $\phi$, which is a function of time.

For an isotropic and homogeneous Friedmann-Robertson-Walker (FRW) space-time, with zero spatial curvature, the line element is given by,

$$ds^2 = c^2dt^2 - a^2(t)[dr^2 + r^2\,d\theta^2 + r^2 sin^2\theta\,d\varphi^2] \tag{2}$$

Where $a(t)$ stands for the scale factor of the universe and the symbols $r$, $\theta$ and $\varphi$ denote the spherical polar coordinates.

Through the variation of the action (eqn. 1), with respect to the components of the metric tensor (eqn. 2), the field equations of the generalized Brans-Dicke theory are obtained as [Banerjee & Ganguly 2009],

$$3\left(\frac{\dot{a}}{a}\right)^2 = \frac{\rho}{\phi} + \frac{\omega(\phi)}{2}\left(\frac{\dot{\phi}}{\phi}\right)^2 - 3\frac{\dot{a}}{a}\frac{\dot{\phi}}{\phi}, \tag{3}$$

$$2\frac{\ddot{a}}{a} + \left(\frac{\dot{a}}{a}\right)^2 = -\frac{\omega(\phi)}{2}\left(\frac{\dot{\phi}}{\phi}\right)^2 - 2\frac{\dot{a}}{a}\frac{\dot{\phi}}{\phi} - \frac{\ddot{\phi}}{\phi}. \tag{4}$$

Equations (3) and (4) represent the characteristics of a matter dominated universe, regarded as a pressureless dust, where the equation-of-state (EoS) parameter ($P/\rho$) is zero.

In these equations, written for $P = 0$, $\rho$ denotes the density of matter (baryonic + dark) [Banerjee & Ganguly 2009].

The wave equation, satisfied by the scalar field ($\phi$), is expressed as,

$$\ddot{\phi} + 3\frac{\dot{a}}{a}\dot{\phi} = \frac{1}{2\omega+3}\left(T - \dot{\phi}^2\frac{d\omega}{d\phi}\right) \tag{5}$$

Where $T$ denotes the trace of the energy-momentum tensor. This equation is not an independent one. It follows from Bianchi identities [Banerjee & Ganguly 2009].

Combining equation (3) with (4), one obtains,

$$2\frac{\ddot{a}}{a} + 4\left(\frac{\dot{a}}{a}\right)^2 = \frac{\rho}{\phi} - 5\frac{\dot{a}}{a}\frac{\dot{\phi}}{\phi} - \frac{\ddot{\phi}}{\phi}. \quad (6)$$

Using equation (3), the coupling parameter $\omega(\phi)$is obtained as,

$$\omega(\phi) = 2\left[3\left(\frac{\dot{a}}{a}\right)^2 - \frac{\rho}{\phi} + 3\frac{\dot{a}}{a}\frac{\dot{\phi}}{\phi}\right]\left(\frac{\dot{\phi}}{\phi}\right)^{-2} \quad (7)$$

## 3 THEORETICAL MODEL

There are theoretical models where the matter content of the universe (dark + baryonic) is assumed to be conserved [Banerjee & Ganguly 2009]. The conservation of matter, although not assumed in the present study, is mathematically represented by the following expression.

$$\rho a^3 = \rho_0 a_0^3 = \rho_0 \qquad \text{(taking } a_0 = 1\text{)} \quad (8)$$

In the framework of Brans-Dicke theory of gravity, there are some studies of cosmology which are based on an interaction between matter and the scalar field [Das & Mamon 2014]. In these studies, one takes into account the possibility of an inter-conversion between dark energy and matter (dark + baryonic). Considering this possibility, we propose the following expression of the density of matter ($\rho$), as a modified form of equation (8).

$$\rho a^3 = f(t)\,\rho_0 a_0^3 = f(t)\,\rho_0 \qquad \text{(taking } a_0 = 1\text{)} \quad (9)$$

The present study is based on a simple observation that the right hand side of equation (8) cannot be independent of time, in a dust universe, if matter does not remain conserved owing to an inter-conversion between dark energy and matter. Here we incorporate a function of time $f(t)$ in equation (8) to obtain equation (9) which clearly represents non conservation of the matter content of universe. This arbitrarily introduced function $f(t) = \frac{\rho a^3}{\rho_0 a_0^3}$, stands for the ratio of the matter content of the universe at any time $t$ to the matter content at the present instant ($t = t_0$). Let us express this ratio as $R_1 \equiv f(t) = M(t)/M(t_0)$ where $M(t) = \rho a^3$ is the matter content of the universe. A second ratio is defined as $R_2 = \frac{1}{f}\frac{df}{dt} = \frac{1}{M}\frac{dM}{dt}$, representing the fractional rate of change of the matter content. A negative value of $R_2$ implies a decay of matter into some other form, resulting from its interaction with dark energy. We have defined a third ratio as $R_3 = f(t) - 1 = \frac{M(t) - M(t_0)}{M(t_0)}$, representing the fractional difference between the matter contents at $t$ and $t_0$, relative to the present content of matter. If $M(t)$ is a monotonically decreasing function of time, one must have $R_3 > 0$ for $t < t_0$ and $R_3 < 0$ for $t > t_0$.

Here we have assumed that a process of transformation of matter into dark energy began in the past at the time of $t = \gamma t_0$ where $\gamma < 1$. Thus, at $t = \gamma t_0$, the total matter content of the universe was $M(\gamma t_0) = M(t_0)R_1(\gamma t_0)$ and the dark energy content can be assumed to be $M_D(\gamma t_0) = \epsilon M(\gamma t_0)$. Here $\epsilon$ is a constant satisfying the condition that $\epsilon < \frac{\Omega_{D0}}{\Omega_{m0}} = \frac{0.7}{0.3} \cong 2.33$, where $\Omega_{D0}$ and $\Omega_{m0}$ are the density parameters of dark energy and matter respectively at the present time, which are defined as $\Omega_{D0} = \rho_{D0}/\rho_{t0}$ and $\Omega_{m0} = \rho_0/\rho_{t0}$ respectively. These symbols, $\rho_0$, $\rho_{D0}$ and $\rho_{t0}$, denote respectively the present values of the matter density, dark energy density and the total matter-energy density. The restriction regarding the value of $\epsilon$ is chiefly due to the consideration that, the change from decelerated expansion to accelerated expansion took place in the past as the dark energy content exceeded the matter content, because, it is the dark energy that is said to have caused the accelerated expansion, as per scientific literature [Banerjee & Ganguly 2009; Goswami 2017]. As the dark energy content has been increasing at the cost of matter, their ratio was definitely less in the past than its value at the present time. In accordance with a recent study [Goswami 2017], this transition took place $7.2371 \times 10^9$ years ago ($z \cong 0.6818$)when the age of the universe was nearly half its present age. Another recent study shows that the universe had $\Omega_m = \Omega_D$ nearly at $z = 0.7$ [Das & Mamon 2014]. Therefore, in the present study, $M(\gamma t_0) + M_D(\gamma t_0) = (1 + \epsilon)M(\gamma t_0)$ is regarded as the total content of matter and dark energy for the entire matter dominated era. Based on these considerations we have defined the following ratios, $R_4$ and $R_5$, denoting the proportions of matter and dark energy of the universe.

$$R_4 = \frac{(1+\epsilon)M(\gamma t_0) - M(t)}{(1+\epsilon)M(\gamma t_0)} = \frac{(1+\epsilon)f(\gamma t_0) - f(t)}{(1+\epsilon)f(\gamma t_0)} = 1 - \frac{f(t)}{(1+\epsilon)f(\gamma t_0)} \qquad \text{with } \gamma < 1 \qquad (10)$$

$$R_5 = 1 - R_4 = \frac{M(t)}{(1+\epsilon)M(\gamma t_0)} = \frac{f(t)}{(1+\epsilon)f(\gamma t_0)} \qquad \text{with } \gamma < 1 \qquad (11)$$

Here, $(R_4 \times 100)$ and $(R_5 \times 100)$ are respectively the percentages of dark energy and matter present in the universe at any time *t*. Approximately 70% of the entire matter-energy content of the universe is dark energy at the present time [Das & Mamon 2014; Pal 2000; Goswami 2017]. The right choice of values of the parameters, $\gamma$, $\epsilon$ and $k$ (to be defined later), will cause $R_4$ and $R_5$ to have values close to those obtained from observations (0.7 and 0.3 respectively).

It has been assumed for the derivations of equations (10) and (11) that the universe is mainly composed of matter and dark energy [Pal 2000; Goswami 2017].

The aim of the present model is to find a functional form of $f(t)$ to determine the time variations of the ratios $R_1, R_2, R_3, R_4$ and $R_5$.

In terms of these parameters, the dark energy density can be expressed as,

$$\rho_D = \frac{R_4}{R_5}\rho = \frac{(1+\epsilon)f(\gamma t_0) - f(t)}{f(t)}\rho \qquad (12)$$

Using equation (12), the density of the entire matter-energy content of the universe becomes,

$$\rho_{total} = \rho_D + \rho = \rho\left(1 + \frac{R_4}{R_5}\right) = \frac{(1+\epsilon)f(\gamma t_0)}{f(t)}\rho \qquad (13)$$

As an expression of $f(t)$ we have used the following ratio ($R_1$) which is based on equation (9).

$$f(t) \equiv R_1 = \frac{M(t)}{M(t_0)} = a^3 \frac{\rho}{\rho_0} \qquad (14)$$

In equation (14), the density of matter ($\rho$) can be incorporated from equation (6). An empirical expression (eqn. 15) of the Brans-Dicke scalar field $\phi$ has been used for this purpose, based on some recent studies [Banerjee & Ganguly 2009; Roy, Chattopadhyay, & Pasqua 2013].

$$\phi = \phi_0 \left(\frac{a}{a_0}\right)^k = \phi_0 a^k \quad \text{(taking } a_0 = 1\text{)} \tag{15}$$

Here, the constant $k$ determines the rate at which the parameter $\phi \left(\equiv \frac{1}{G}\right)$ changes with time. Using equation (15) in (6), the density of matter ($\rho$) is obtained as,

$$\rho = \phi H^2[k^2 + (4-q)k + (4-2q)] \tag{16}$$

We have used the standard expressions of the Hubble parameter ($H$) and deceleration parameter ($q$), which are $H = \dot{a}/a$ and $q = -\ddot{a}a/\dot{a}^2$ respectively, in deriving equation (16). We have obtained equation (16) by substituting equation (15) into equation (6), which was obtained from Brans-Dicke field equations. Since equation (16) has its origin in the field equations for a spatially flat matter-dominated universe, the nature of its time dependence is supposed to be qualitatively correct, but the present value of matter density ($\rho_0$) obtained from it may not be the same as the value of $\rho_0$ obtained from astrophysical observations, owing to the arbitrariness in assuming the ansatz for $\varphi$ (eqn. 15). This ambiguity can be rectified by defining the matter density ($\rho_m$) in the following way.

$$\rho_m = \rho_0 \frac{\phi H^2[k^2+(4-q)k+(4-2q)]}{[\phi H^2[k^2+(4-q)k+(4-2q)]]_{t=t_0}} \tag{17}$$

We have obtained equation (17) from equation (16) only by multiplying the latter with a suitable constant. In equation (17), the present density of matter of the universe has been taken to be $\rho_0 = 2.831 \times 10^{-27} Kg/m^3$, in accordance with WMAP data [Goswami 2017; Pradhan, Amirhashchi & Saha 2011]. From (17), one gets $\rho_m = \rho_0$ at $t = t_0$.
Equations (12) and (13) should be modified so that one gets correct values of $\rho_D$ and $\rho_{total}$, respectively, from them at $t = t_0$. In this regard one must take $\rho = \rho_m$ in those equations, where $\rho_m$ is expressed by equation (17).
Taking $\rho = \rho_m$ in equation (14) and then using (17) in it, one gets,

$$f(t) = a^3 \frac{\phi H^2[k^2+(4-q)k+(4-2q)]}{\phi_0 {H_0}^2[k^2+(4-q_0)k+(4-2q_0)]} = a^3 \frac{\phi H^2}{\phi_0 {H_0}^2} \frac{(k+2-q)}{(k+2-q_0)} \tag{18}$$

The parameters $\phi$, $H$ and $q$, in equation (18), are time dependent quantities. Their time variations depend upon the time evolution of the scale factor ($a$) from which they are to be determined. An empirical scale factor has been used to evaluate $f(t)$ from equation (18). This empirical form has been chosen to be consistent with a fact regarding the deceleration parameter. It is based on observations that the universe had a phase of decelerated expansion before the beginning of its present phase of accelerated expansion [Das & Mamon 2014; Banerjee & Das 2006; Banerjee & Ganguly 2009]. A signature flip of the deceleration parameter must have occurred due to this change of phase. Our scale factor has a functional form which ensures this change of sign of the deceleration parameter. This function is a product of exponential function and power law, which was used by Roy *et al.* and Pradhan *et al.* [Roy, Chattopadhyay & Pasqua 2013; Pradhan, Saha & Rikhvitsky 2015]. It can be written as,

$$a = a_0 \ (t/t_0)^\alpha \ Exp[\beta(t - t_0)] \tag{19}$$

In equation (19), we must have $\alpha, \beta > 0$ to ensure an increase of scale factor with time. The Brans-Dicke scalar field ($\phi$), Hubble parameter ($H$) and the deceleration parameter ($q$), based on this scale factor are expressed as,

$$\phi = \phi_0 \left(\frac{a}{a_0}\right)^k = \phi_0 \ (t/t_0)^{k\alpha} \ Exp[\beta k(t - t_0)] \tag{20}$$

$$H = \dot{a}/a = \beta + \frac{\alpha}{t} \tag{21}$$

$$q = -\ddot{a}a/\dot{a}^2 = -1 + \frac{\alpha}{(\alpha+\beta t)^2} \tag{22}$$

For $0 < \alpha < 1$, one gets $q > 0$ at $t = 0$ and, for $t \to \infty$, one gets $q \to -1$.
It shows a clear change of sign of the scale factor with time.
Following conditions have been used to determine the values of the constant parameters $(\alpha, \beta)$.

Condition 1: $H = H_0$ at $t = t_0$ (23)
Condition 2: $q = q_0$ at $t = t_0$ (24)

Using equations (23) and (24), in (21) and (22) respectively, one gets,

$$\alpha = (1 + q_0)(H_0 t_0)^2 = 4.76 \times 10^{-01} \tag{25}$$
$$\beta = \frac{H_0 t_0 - \alpha}{t_0} = \frac{H_0 t_0 - (1+q_0)(H_0 t_0)^2}{t_0} = 1.25 \times 10^{-18} \tag{26}$$

For the present study we have used the following values of cosmological parameters.

$$H_0 = 72 \left(\frac{Km}{Sec}\right) per\ Mega\ Parsec\ = 2.33 \times 10^{-18} sec^{-1} = 7.35 \times 10^{-11}\ Yr^{-1}$$
$$t_0 = 14\ billion\ years = 4.415 \times 10^{17} sec$$
$$\varphi_0 = \frac{1}{G_0} = 1.498 \times 10^{10} m^{-3} Kg s^2$$
$\rho_0 = 2.831 \times 10^{-27} Kg/m^3$ [present density of matter (dark+baryonic)]
$$q_0 = -0.55$$

To obtain the time variation of $f(t)$, from equation (18), one has to use equations (19), (20), (21), (22), (25), (26) and also the above mentioned values of cosmological parameters.
One can determine the time evolution of $\rho_m$ by using equations (20), (21), (22), (25) and (26) in equation (17).
Equation (18) can be used to determine the time dependence of $R_3[\equiv f(t) - 1]$.
The time dependence of the parameters $R_4$ and $R_5$ can be studied from equations (10) and (11) respectively by combining equation (18) with them.

The expression of $R_2$, from equations (18), (19), (20), (21) and (22), is given by,

$$R_2 = \frac{1}{f}\frac{df}{dt} = \beta(3 + k) + \frac{\alpha(3+k)-2}{t} + \frac{2\beta(3+k)(\alpha+\beta t)}{\alpha[\alpha(3+k)-1]+2\alpha\beta(3+k)t+\beta^2 t^2(3+k)} \tag{27}$$

To estimate the time variation of $\rho_D$, $\rho_{total}$, using equations (12) and (13) respectively, one must use $\rho = \rho_m$ in these equations. Then, one has to substitute equation (17) into them and use equation (18), for the determination of $f(t)$ and $f(\gamma t_0)$.
The defining relation of the function $f(t)$ is $\rho a^3 = f(t)\, \rho_0 a_0^3$. It shows that the value of $f(t)$ must be positive and, we must have, $f(t) = 1$ at $t = t_0$ (taking $a_0 = 1$). The form $f(t)$, as per equation (18), ensures that $f(t) = 1$ at $t = t_0$. The ranges of the parameter $k$, for which $f(t) > 0$, throughout the range of study (say, from $t = 0.5t_0$ to $t = 1.5t_0$) are given below.

$k < k_1$ and $k > k_2$ where,

$$k_1 = (q-2)_{min} = q_{(t=1.5t_0)} - 2 \tag{28}$$

$$k_2 = (q-2)_{max} = q_{(t=0.5t_0)} - 2 \tag{29}$$

Using $q(t)$ from equation (22) we get $k_1 = -2.73$ and $k_2 = -2.15$

Therefore, the lower and the upper ranges of permissible values for $k$ are $k < k_1$ and $k > k_2$ respectively. The range of $k > k_2$ has both positive and negative values of $k$ and the range of $k < k_1$ has only negative values. For negative values of $k$, equation (15) shows the parameter $\phi$ to be a decreasing function of time, implying an increase of the gravitational constant $(G \equiv \frac{1}{\phi})$ with time. Thus, the range of $k > k_2$ causes $G$ to be both increasing and decreasing functions of time, and the range of $k < k_1$ causes $G$ to be an increasing function only. Some recent studies have shown the gravitational constant to be increasing with time [Pradhan, Saha, & Rikhvitsky 2015; Saha, Rikhvitsky & Pradhan 2015]. One gets a valid reason for choosing negative values of $k$ from these studies.
To choose the correct range of $k$ values, we have also calculated $\omega_0$ (the value of BD parameter at the present time) for different values of $k$ and compare them with the findings of other studies. Substituting (15) into (7), we get the following functional form of $\omega$.

$$\omega = \frac{2}{k^2}\left[3(1+k) - \frac{\rho}{\phi H^2}\right] \tag{30}$$

To obtain the time variation of $\omega$ from equation (30), we have to use equations (16), (20) and (21) for the values of $\rho$, $\phi$ and $H$.
One gets the following expression of $\omega_0$ from equations (30).

$$\omega_0 = \frac{2}{k^2}\left[3(1+k) - \frac{\rho_0}{\phi_0 {H_0}^2}\right] \tag{31}$$

According to several studies on Brans-Dicke theory, $\omega_0$ has a small negative value [Banerjee & Ganguly 2009; Sahoo & Singh 2002].
To get $\omega_0 < 0$ (based on equation 31), $k$ must satisfy the following conditions.

$$k < \frac{\rho_0}{3\, \phi_0 {H_0}^2} - 1 \quad \text{or,} \quad k < -0.99 \tag{32}$$

Over a part of the upper range of $k$ values (i.e. $k > k_2$) and all over its lower range (i.e. $k < k_1$), equation (32) would be satisfied.

The following expression of the gravitational constant ($G$), which is the reciprocal of the Brans-Dicke scalar field ($\phi$), has been obtained from equation (20).

$G = \frac{1}{\phi} = \frac{(a/a_0)^{-k}}{\varphi_0} = \frac{1}{\varphi_0} Exp\left[k\alpha t_0^{\beta}\right] Exp[-k\alpha t^{\beta}]$

Thus, $\frac{G}{G_0} = Exp\left[k\alpha t_0^{\beta}\right] Exp[-k\alpha t^{\beta}]$ (33)

Substituting $\phi = \frac{1}{G}$ and $\phi_0 = \frac{1}{G_0}$ , in equation (18), one gets,

$$\frac{G}{G_0} = a^3 \frac{H^2}{{H_0}^2} \frac{(k+2-q)}{(k+2-q_0)} \frac{1}{f(t)} \quad (34)$$

Here, $\frac{G}{G_0}$ has been expressed in terms of the function $(t)$ $(\equiv R_1)$. The parameters $H$ and $q$ in this expression are functions of the scale factor $(a)$. Any decrease of $f(t)$ will have an increasing effect on $\frac{G}{G_0}$. From equation (34) we get the relation $\frac{\partial G}{\partial f} = -\frac{G}{f}$ which is negative since $G$ and $f$ are both positive, and it becomes more negative with time since $G$ and $f$ are respectively increasing and decreasing functions of time (as shown in figures 3 and 10 respectively).

A quantity $\frac{\dot{G}}{G}$ which can be determined from astrophysical observations, is given by,

$$\frac{\dot{G}}{G} = \frac{1}{G}\frac{dG}{dt} = -k\frac{\dot{a}}{a} = -kH = -k\left(\beta + \frac{\alpha}{t}\right) \quad (35)$$

Time evolutions of $\frac{G}{G_0}$ and $\frac{\dot{G}}{G}$ can be determined from equations (33) and (35) respectively.
Using equation (35) one gets,

$$\left(\frac{\dot{G}}{G}\right)_{t=t_0} = -kH_0 \quad (\text{where } H_0 = 7.35 \times 10^{-11}\, Yr^{-1}) \quad (36)$$

The value of the parameter $k$ should such that $\left|\frac{\dot{G}}{G}\right|_{t=t_0} < 4 \times 10^{-10}\ Yr^{-1}$ [Weinberg 1972].
Therefore we have, $k > -5.44$.

Thus, from all these discussions, the range of variation of the parameter $k$ is found to be $-5.44 < k < -2.73$ and $-2.15 < k < -0.99$ which ensures that $\omega_0 < 0$ and $\left|\frac{\dot{G}}{G}\right|$ remains less than its permissible upper limit, as obtained by S. Weinberg [Weinberg 1972].
Hence, the lower bound of the lower range of $k$ $(i.e. k < k_1)$ and the upper bound of the upper range of $k$ $(i.e. k > k_2)$ are $-5.44$ and $-0.99$ respectively.
On the basis of these findings we have found it logical to use only negative values of the parameter $k$ to determine the time evolution of all cosmological quantities that depend on its value in the present model.

# 4 DETERMINATION OF $f(t)$ FROM THE INTERACTION OF MATTER AND DARK ENERGY

The differential equation, representing the conservation of the entire matter-energy content of the universe is expressed in the following way [Sahoo & Singh 2002].

$$\dot{\rho}_{total} + 3H(\rho_{total} + P) = 0 \tag{37}$$

Matter and dark energy are the major constituents of the matter dominated present era of the universe [Pal 2000; Goswami 2017]. Therefore we have $\rho_{total} = \rho + \rho_D$ where $\rho$ and $\rho_D$ denote respectively the densities of matter and dark energy.
The pressure $(P)$ of all constituents of the universe is contributed by dark energy because, the entire content of matter (dark matter + baryonic matter) is regarded as pressureless dust [Banerjee & Ganguly 2009; Farajollahi & Mohamadi 2010]. Hence we write,

$$P = \gamma\rho_{total} = \gamma_D\rho_D \tag{38}$$

In equation (38), $\gamma$ and $\gamma_D$ denote respectively the equation-of-state (EoS) parameters for total energy and dark energy.
Combining equation (38) with (37) and using $\rho_{total} = \rho + \rho_D$ we get,

$$\dot{\rho} + \dot{\rho}_D + 3H[\rho + \rho_D(1 + \gamma_D)] = 0 \tag{39}$$

Assuming the two constituents, matter and dark energy, to be interacting with each other, and generating one at the cost of the other, it is possible to define a parameter representing their interaction, using equation (39). This interaction term $(Q)$ is expressed by the following equations [Das & Mamon 2014; Farajollahi & Mohamadi 2010; Abdollahi Zadeh, Sheykhi & Moradpour 2017].

$$\dot{\rho} + 3H\rho = Q \tag{40}$$

$$\dot{\rho}_D + 3H\rho_D(1 + \gamma_D) = -Q \tag{41}$$

Transfer of energy from the field of matter to the sector of dark energy is indicated by a negative value of $Q$. A positive value of $Q$ implies transformation of dark energy into matter.
In order to express $f(t)$ as a function of the matter-energy interaction term $(Q)$, we have formulated the following two models.

## 4.1 Model – 1

In this model, the following ansatz is assumed to solve equation (40).

$$Q = \lambda_1 H \tag{42}$$

Here, the constant $\lambda_1$ has the dimension of $\rho$.
Substituting equation (42) into (40) and using the expression of $H$ (eqn. 21), we get the following expression of density.

$$\rho = \frac{\lambda_1}{3} - \frac{1}{3}\left[\beta(t - t_0) + \alpha\, ln\left(\frac{t}{t_0}\right) + (\lambda_1 - 3\rho_0)^{-1/3}\right]^{-3} \tag{43}$$

Using equation (43) in (14) we get,

$$f(t) = a^3 \frac{\rho}{\rho_0} = \frac{\lambda_1 a^3}{3\rho_0} - \frac{a^3}{3\rho_0}\left[\beta(t - t_0) + \alpha\, ln\left(\frac{t}{t_0}\right) + (\lambda_1 - 3\rho_0)^{-1/3}\right]^{-3} \qquad (44)$$

For any value of $\lambda_1$ in equation (44), $f(t_0) = {a_0}^3 = 1$. This is a condition to be satisfied by $f(t)$ according to its definition obtained from equation (14).
Using equation (42), equation (44) takes the following form.

$$f(t) = a^3 \frac{\rho}{\rho_0} = \frac{a^3}{3\rho_0}\frac{Q}{H} - \frac{a^3}{3\rho_0}\left[\beta(t - t_0) + \alpha\, ln\left(\frac{t}{t_0}\right) + \left(\frac{Q}{H} - 3\rho_0\right)^{-1/3}\right]^{-3} \qquad (45)$$

In equation (45), $f(t)$ is expressed in terms of the matter-energy interaction term ($Q$).

### 4.2 Model – 2

In this model, the following ansatz is assumed to solve equation (40).

$$Q = \lambda_2 \rho \qquad (46)$$

Here, the constant $\lambda_2$ has the dimension of the Hubble parameter.

Using equation (46) in equation (40) and using the expression of $H$ (eqn. 21), we get the following expression for density.

$$\rho = \rho_0 Exp\left[\lambda_2(t - t_0) - 3\left\{\beta(t - t_0) + \alpha\, ln\left(\frac{t}{t_0}\right)\right\}\right] \qquad (47)$$

Using equation (47) in (14) we get,

$$f(t) = a^3 \frac{\rho}{\rho_0} = a^3 Exp\left[\lambda_2(t - t_0) - 3\left\{\beta(t - t_0) + \alpha\, ln\left(\frac{t}{t_0}\right)\right\}\right] \qquad (48)$$

In equation (48), for any value of $\lambda_2$, we have $f(t_0) = {a_0}^3 = 1$. This is a condition to be satisfied by $f(t)$, as per its definition obtained from equation (14).
Using equation (46), equation (48) takes the following form.

$$f(t) = a^3 Exp\left[\frac{Q}{\rho}(t - t_0) - 3\left\{\beta(t - t_0) + \alpha\, ln\left(\frac{t}{t_0}\right)\right\}\right] \qquad (49)$$

In equation (49), $f(t)$ is expressed in terms of the matter-energy interaction term ($Q$).

Combining equation (18) with (48), we get the following expression of $\lambda_2$.

$$\lambda_2 = (t - t_0)^{-1}\left[ln\frac{\varphi H^2[k^2 + (4 - q)k + (4 - 2q)]}{\varphi_0 {H_0}^2[k^2 + (4 - q_0)k + (4 - 2q_0)]} + 3\left\{\beta(t - t_0) + \alpha\, ln\left(\frac{t}{t_0}\right)\right\}\right] \qquad (50)$$

From the equations (44) and (48), one obtains the following relationship between $\lambda_1$ and $\lambda_2$.

$$\frac{\lambda_1 a^3}{3\rho_0} - \frac{a^3}{3\rho_0}\left[\beta(t-t_0) + \alpha\, ln\left(\frac{t}{t_0}\right) + (\lambda_1 - 3\rho_0)^{-\frac{1}{3}}\right]^{-3}$$
$$= a^3 Exp\left[\lambda_2(t-t_0) - 3\left\{\beta(t-t_0) + \alpha\, ln\left(\frac{t}{t_0}\right)\right\}\right] \tag{51}$$

Using equation (49), the following expression for the interaction term ($Q$), between matter and dark energy, is obtained.

$$Q = \frac{\rho}{t-t_0}\left[3\left\{\beta(t-t_0) + \alpha\, ln\left(\frac{t}{t_0}\right)\right\} + ln\frac{f(t)}{a^3}\right] \tag{52}$$

To estimate the time dependence of $Q$ from equation (52), one should use equations (16), (18) and (19) for $\rho$, $f(t)$ and $a$ respectively. From equation (52), one gets $Q = -1.484 \times 10^{-44}$ at $\frac{t}{t_0} = 0.998$ and $Q = -1.450 \times 10^{-44}$ at $\frac{t}{t_0} = 1.002$. Its negative sign implies a decay of matter into dark energy. As time goes on, $|Q|$ becomes closer to zero, indicating a gradual reduction in matter-energy interaction.
From equations (42) and (46), we get $\lambda_1 = -6.4 \times 10^{-27}$ and $\lambda_2 = -5.3 \times 10^{-18}$ respectively, taking $Q = -1.5 \times 10^{-44}$ at $t = t_0$.
Using equation (41), we get the following expression of the equation-of-state (EoS) parameter for dark energy ($\gamma_D$).

$$\gamma_D = -1 - \frac{Q + \dot{\rho}_D}{3H\rho_D} \tag{53}$$

To find the time variation of the EoS parameter ($\gamma_D$), based on equation (53), one should use equations (12), (21) and (52) to get values of $\rho_D$, $H$ and $Q$ respectively.

Using the relation $R_2 = \frac{1}{f}\frac{df}{dt}$ and equations (9), (12), (53) and one gets,

$$\gamma_D = -1 - \frac{Q}{3H\rho_D} - \frac{1}{3H\rho_D}\left[\frac{\rho_0 f}{a^3}(R_2 - 3H)\left\{\frac{(1+\epsilon)\, f(\gamma t_0)}{f} - 1\right\} - \frac{(1+\epsilon)\, f(\gamma t_0)\rho R_2}{f}\right] \tag{54}$$

For the derivation of equation (54), we have used the relations $\dot{\rho} = \frac{\rho_0 f}{a^3}(R_2 - 3H)$ and $\dot{\rho}_D = \left[\dot{\rho}\left\{\frac{(1+\epsilon)\, f(\gamma t_0)}{f} - 1\right\} - \frac{(1+\epsilon)\, f(\gamma t_0)\rho R_2}{f}\right]$, formulated from equations (9) and (12) respectively, using the relations $\dot{a} = aH$ and $\dot{f} = fR_2$ for them.
One can determine the time dependence of the EoS parameter ($\gamma_D$) by using equations (12), (16), (18), (19), (21), (27) and (52) in (54). Equation (54) expresses $\gamma_D$ as a function of $f(t)$, which is proportional to the matter content of the universe.

The approximate range of variation of the equation-of-state (EoS) parameter ($\gamma_D$), for dark energy, is found to be $-1.1 \le \gamma_D \le -0.9$, as obtained from some recent studies in this regard [Wood-Vasey et al. 2007; Davis et al. 2007]. Three sets of values of the parameters discussed in the present study have been shown in Table-1. The values of these parameters ($\gamma, \epsilon, k$) have been so chosen that the values of the dependent ones fall within their permissible limits. In Table-1, $R_{40}$, $R_{50}$ and $\gamma_{D0}$ denote the values of $R_4$, $R_5$ and $\gamma_D$ at the present time ($t = t_0$). One may choose several such sets of values under the constraints of $\gamma < 1$, $\epsilon < 2.33$, $-5.44 < k < -2.73$ and $-2.15 < k < -0.99$ , as obtained earlier in this article.

| **TABLE – 1** Different sets of parameter values obtained from the present model | | | | | | | |
|---|---|---|---|---|---|---|---|
| | Independent Parameters | | | Cosmological Quantities | | | |
| | $\gamma$ | $\epsilon$ | $k$ | $\omega_0$ | $R_{40}$ | $R_{50}$ | $\gamma_{D0}$ |
| Set-1 | 0.580 | 0.000 | -3.900 | -1.149 | 0.705 | 0.295 | -1.119 |
| Set-2 | 0.600 | 0.100 | -3.930 | -1.143 | 0.713 | 0.287 | -1.069 |
| Set-3 | 0.620 | 0.165 | -3.895 | -1.150 | 0.705 | 0.295 | -1.001 |

# 5 GRAPHICAL DEPICTION AND INTERPRETATION OF THE THEORETICAL FINDINGS

In Figure 1, $\omega_0$ has been plotted as a function of $k$. For the lower range of the values of $k$ ($k < k_1$), the values of $\omega_0$ are negative and also close to the values obtained from other studies [Banerjee & Pavon 2001; Sahoo & Singh 2002]. For the upper range of the values where $k > k_2$, $\omega_0$ can be both positive and negative. Its positive values are in complete disagreement with the findings of other studies on generalized BD theory, where $\omega$ is not regarded as a constant [Banerjee & Ganguly 2009; Sahoo & Singh 2002].

For the plot of $\omega_0$ versus $k$, in Fig. 1, we have used a set of data (based on eqn. 31), that does not contain the data of $k = 0$. A smooth curve has been drawn connecting these data points. The value of $k = 0$ is physically unacceptable, because it implies (as per eqn. 15) that the Brans-Dicke scalar field ($\phi$) is independent of time, which is inconsistent with the Brans-Dicke theory of gravity [Brans & Dicke 1961]. Mathematically, $\omega_0$ would be positive infinity at $k = 0$, according to equation (31). Since $k$ cannot ever be zero, the value of $\omega_0$ at $k = 0$, obtained from this graph, is absolutely of no interest, validity and relevance. The values of $\omega_0$, only for non-zero values of $k$, are supposed to have physical relevance or importance.

The gravitational constant increases with time for negative values of $k$, according to equation (35). Thus, for $k < k_1$, $G$ increases with time. For a very small part of the range of $k > k_2$, $G$ is an increasing function of time. There are theoretical models, based on astrophysical observations, where the gravitational constant has been shown to be increasing with time [Pradhan, Saha & Rikhvitsky 2015; Saha, Rikhvitsky & Pradhan 2015].

We have shown the time variation of the Brans-Dicke parameter $\omega$ in figure-2 for a value of $k$ in its lower range and also for a negative value in its upper range. The first one is an increasing function of time, becoming less negative with time, and this behavior is sufficiently consistent with other studies [Sahoo & Singh 2002]. The behaviour of the second one is quite the contrary. For the positive values of the upper range of $k$ (i.e. $k > k_2$), the values of $\omega$ are positive, as given by equation (30), and they are not at all consistent with the results of other studies [Banerjee & Ganguly 2009].

In accordance with some earlier studies, the value of $\omega_0$ should be lying in the range of $-2 < \omega_0 < -1$ [Banerjee & Pavon 2001; Bertolami & Martins 2000; Sen & Seshadri 2003; Sahoo & Singh 2003]. The values of $\omega_0$ that we get from this model, for the lower range of $k$ values (i.e. $k < k_1$), are consistent with this result.

Based on the observations of the last four paragraphs, it would be reasonable to use the values of $k$ from its lower range (i.e. $k < k_1$) to determine the time evolution of $f(t)$ and other relevant cosmological parameters connected to it in the present study.

Figure 3 depicts the time variation of $R_1 (\equiv f)$ for three different values of $k$ chosen from its lower range. It is evident that the matter content of the universe $[M(t) = f(t)M_0]$ decreases with time at a rate which is greater for more negative values of $k$.

Figure 4 shows the plots of $R_2 \left(\equiv \dot{f}/f\right)$ as a function of time for three different values of $k$ in its lower range. It becomes less negative with time at a gradually decreasing rate. At any value of $(t/t_0)$, $R_2$ is larger for less negative values of the parameter $k$.

The time variation of $R_3$ has been depicted in Figure 5 for three different values of $k$ in the range of $k < k_1$. Here, $R_3 \left(\equiv f(t) - 1 = \frac{M(t)-M(t_0)}{M(t_0)}\right)$ decreases with time at a gradually decreasing rate. As expected from its expression, it is zero at the present time. One finds it to be positive for $t < t_0$ and, it becomes negative for $t > t_0$, showing clearly a reduction in the matter content of the universe $[M(t)]$ with time. It is evident that more negative values of the parameter $k$ causes faster decrease of $R_3$ with time.

Figure 6 shows the variation of $R_2$ and $R_3$ as functions of $R_1 (\equiv f)$. It is evident from Figure 3 that $R_1$ decreases with time monotonically. By definition, the product of $R_1$ and $R_2$ is proportional to the rate of change of $R_1$. As $R_1$ approaches its present value of unity, $R_2$ becomes less negative, indicating a slower rate of decay of matter into dark energy. Here we find $R_3$ to be positive in the past ($R_1 > 1$) and negative in future ($R_1 < 1$), as expected from its definition and the characteristics of $R_1$.

Figure 7 shows the plots of $R_4$ and $R_5$ as functions of time. As time goes on, the proportion of dark energy ($R_4$) of the universe increases with time. Since we assume it to be generated from matter (dark + baryonic), the proportion of matter in the universe ($R_5$) decreases with time. Therefore, the sum of $R_4$ and $R_5$ remains unity because they are the main constituents of the universe at its present phase of expansion [Pal 2000; Goswami 2017]. For $k = -3.87$, $\gamma = 0.6$ and $\epsilon = 0.1$, these curves approximately give $R_4 = 0.7$ and $R_5 = 0.3$ at the present epoch (i.e. $t = t_0$), which are quite consistent with present observations [Pal 2000; Goswami 2017]. Several other combinations of the parameters $k$, $\gamma$ and $\epsilon$ are possible for which one gets correct values of $R_4$ and $R_5$ at $t = t_0$.

Figure 8 depicts the time variations of the density of dark energy ($\rho_D$), density of total matter-energy content ($\rho_{total}$) and the density of matter ($\rho_m$). We have used equations (12), (13) and (17) for these plots. Taking $k = -3.87$, $\gamma = 0.6$ and $\epsilon = 0.1$, the values obtained from these curves at $t = t_0$, are in agreement with the present observations [Goswami 2017]. The plot of time evolution of $\rho_{total}$ is similar to the plots of energy density versus time in other recent studies [Pradhan 2013; Yadav, Rahaman & Ray 2011].

Plots of $H$ and $q$ versus $R_1$ have been shown in figure-9. Data for this plot have been obtained from the time dependent expressions of these three parameters. As $R_1$ moves closer its present value of unity, the dark energy content of universe increases although its rate of generation from matter decreases with time. From these plots one may infer that a larger proportion of dark energy causes faster changes in Hubble parameter and deceleration constant.

Figure 10 depicts the plots of $G/G_0$ as a function of time for three values of $k$ in its lower range ($k < k_1$). It is found to increase with time with a gradually increasing slope. Its rate of increase becomes faster for more negative values of the parameter $k$.

Figure 11 shows the time variation of $\dot{G}/G$ for three values of $k$ in its lower range ($k < k_1$). It is found to be positive and it decreases with time at a gradually decreasing rate. At all values of $(t/t_0)$, it has greater values for more negative values of the parameter $k$.

The variations of $G/G_0$ and $\dot{G}/G$ as functions of $R_1$ have been shown in figure-12. As $R_1$ gets closer to its present value of unity, the dark energy of universe increases although its rate of generation from matter decreases with time. It is evident from these curves that, gravitational constant increases, as dark energy increases. It may also be inferred that, as the rate of generation of dark energy from matter becomes slower, the values of $\dot{G}/G$ decreases. As per figure-3, more negative values of the parameter $k$ causes faster change of $R_1$, indicating a faster generation of dark energy. One finds that for any fixed value of $k$, $R_1$ decreases with time at a gradually decreasing rate, showing a slower rise in dark energy. Hence, the content of dark energy and its rate of production, at the cost of matter, may be regarded to have a role in controlling the behavior of several cosmological parameters connected to the expansion of the universe.

The time dependence of the interaction term ($Q$) between matter and dark energy has been shown in figure-13. Negative value of $Q$ implies a transfer of energy from the sector of matter to dark energy. It becomes less negative with time, becoming asymptotic to zero, implying a gradual reduction in the degree of interaction between matter and dark energy.

The variation of the interaction term ($Q$), as a function of $R_1[\equiv f(t)]$, has been shown in figure-14. By definition, $R_1$ is proportional to the present matter content of the universe and it decrease with time at a gradually decreasing rate (as evident from figure 3). This graph leads one to conclude that $|Q|$ decreases as the matter content of the universe becomes smaller. As $R_1$ gets closer to its present value of unity, at a gradually decreasing rate, $|Q|$ also becomes smaller with time at a gradually slower pace.

Time variation of the equation of state (EoS) parameter for dark energy ($\gamma_D$) has been shown in figure-15. Over a span of time close to the present epoch ($t = t_0$), it has negative values and it gradually approaches zero with time. This feature is reasonably consistent with the findings of other studies of cosmology, in the framework of an anisotropic space-time, based on Einstein's theory of general relativity [Adhav, Bansod & Wankhade 2011; Katore & Shaikh 2015; Pradhan, Amirhashchi & Saha 2011; Vinutha et al. 2016; Yadav, Rahaman & Ray 2011].

According to equation (12), we get $\rho_D \geq 0$ (for all possible values of $\epsilon$), only over a span of time beginning at $t = \gamma t_0$. This is consistent with the fact that the matter content of the universe started getting transformed into dark energy at $t = \gamma t_0$. Any negative value of $\rho_D$ is physically unacceptable. In Figure-15, therefore, the time dependence of $\gamma_D$ has been shown over a period of time starting from $t = \gamma t_0$.

The variation of the equation of state (EoS) parameter for dark energy ($\gamma_D$), as a function of $R_1[\equiv f(t)]$, has been shown in figure-16. The parameter $R_1$ is known to decrease with time (figure 3) and $|Q|$ is known to decrease as $R_1$ decreases (figure 14). It may be inferred that, as the matter content decreases along with a decrease of the rate of conversion from matter to dark energy, the rate of change of the EoS parameter becomes slower (during a period close to the present epoch).

It is found from equation (12) that $\rho_D \geq 0$, for all possible values of the parameter $\epsilon$, only over the span of time during which we have $f(t) \leq f(\gamma t_0)$. This span of time begins at the instant of $t = \gamma t_0$, since $f(t)$ has been found to be a decreasing function of time. This is consistent with the fact that the matter content of the universe started getting transformed into dark energy at $t = \gamma t_0$. Any negative value of $\rho_D$ is physically unacceptable. Therefore, in Figure 16, the dependence of $\gamma_D$ upon $R_1$ $[\equiv f(t)]$ has been shown over a range of $R_1$ values where $R_1 \leq f(\gamma t_0)$.

# 6 CONCLUSIONS

This article is based mainly on the premises of a possible non-conservation of the matter content ($\rho a^3$) in a universe filled with pressureless dust, assuming it to be connected to the decay of matter into dark energy. Here we have assumed that the density of matter ($\rho$) can be expressed in terms of an arbitrarily introduced function of time $f(t)$ (eqn. 9), which is proportional to the matter content of the universe at any instant of time $t$. No assumption has been made regarding the functional form of $f(t)$. We have determined its expression from the field equations of the BD theory using a widely used expression of the scalar field parameter ($\phi$) in terms of the scale factor ($a$). The time dependence of the scale factor has been chosen empirically so that it leads to a deceleration parameter ($q$) that changes sign from positive to negative, as time goes on, implying a transition from deceleration to acceleration of the expanding universe. An important finding of this study is that the function $f(t)$ decreases monotonically with time, indicating a conversion of the matter into dark energy. We have calculated the time variations of the proportions of matter and dark energy contents of the universe in terms of the function $f(t)$. Density of dark energy ($\rho_D$) and the density of the entire matter-energy content ($\rho_{total}$) of universe have been obtained. In the present model, the interaction term between matter and dark energy ($Q$) has been obtained in terms of $f(t)$. An expression of the equation of state (EoS) parameter for dark energy, as a function of $f(t)$, has been derived. The graphical representations of theoretical findings show that, with the rise of the dark energy content of the universe, the gravitational constant ($G$) and the Hubble parameter ($H$) increases and decreases respectively and the deceleration parameter ($q$) continues to be more negative. These findings imply that the time variations of the these cosmological parameters are dependent upon the dark energy content of the universe and its rate of generation owing to an interaction between matter and dark energy, that is decreasing gradually with time, causing $|Q|$ to decrease. The present study shows that if the dark energy is assumed to be generated only from matter, the present proportions of these two components must depend on the duration of their interaction and inter-conversion. By a proper tuning of parameters ($k, \gamma, \epsilon$), the values of the present proportions of matter and dark energy, obtained from this model, become consistent with observations. Through an analysis and comparison of the graphs of the present paper, we can estimate the dependence of cosmological parameters upon the time evolution of matter and dark energy. The energy density ($\rho_{total}$), the EoS parameter for dark energy ($\gamma_D$) and the matter-energy interaction term $(Q)$ have been plotted as functions of time. The characteristics of their time variations are consistent with the findings of other recent studies in this regard [Pradhan 2013; Pradhan, Amirhashchi & Saha 2011; Cueva & Nucamendi 2010]. This consistency of these results with the findings of other recent studies, based on General Relativity or Brans-Dicke theory of gravitation, implies the correctness of the definition and derivation of the function $f(t)$ and the derivations based upon it. A simple theoretical method is provided by this model enabling us to determine the time evolution of both matter and dark energy. In the Figures 7, 8, 15 and 16, showing the variations of cosmological quantities connected to dark energy, the ranges of the independent variables have been chosen in a manner such that $f(t) \leq f(\gamma t_0)$, only to ensure that $\rho_D \geq 0$ (for all values of the parameter $\epsilon$), in accordance with equation (12). This condition is satisfied during the period of time that starts at $t = \gamma t_0$. It is quite consistent with the fact that the matter content of the universe started decaying into dark energy at $t = \gamma t_0$. As a future plan to improve this model, one may think of formulating a theoretical scheme to derive a better form of the function $f(t)$, such that we have $\rho_D \geq 0$ at all values of the cosmic time $(t)$. This model can also be improved by assuming an ansatz for the scalar field $(\phi)$ in a form that is different from equation (15). An empirical expression of the scale factor $(a)$, different from the one given by equation (19), can be used to determine time dependence of several cosmological quantities.

**Acknowledgements** I would like to express my heartfelt gratitude to all academicians and researchers whose works have inspired me to carry out the present study. I am also very thankful to all my colleagues and students for the cooperation I receive at our institution.

# FIGURES

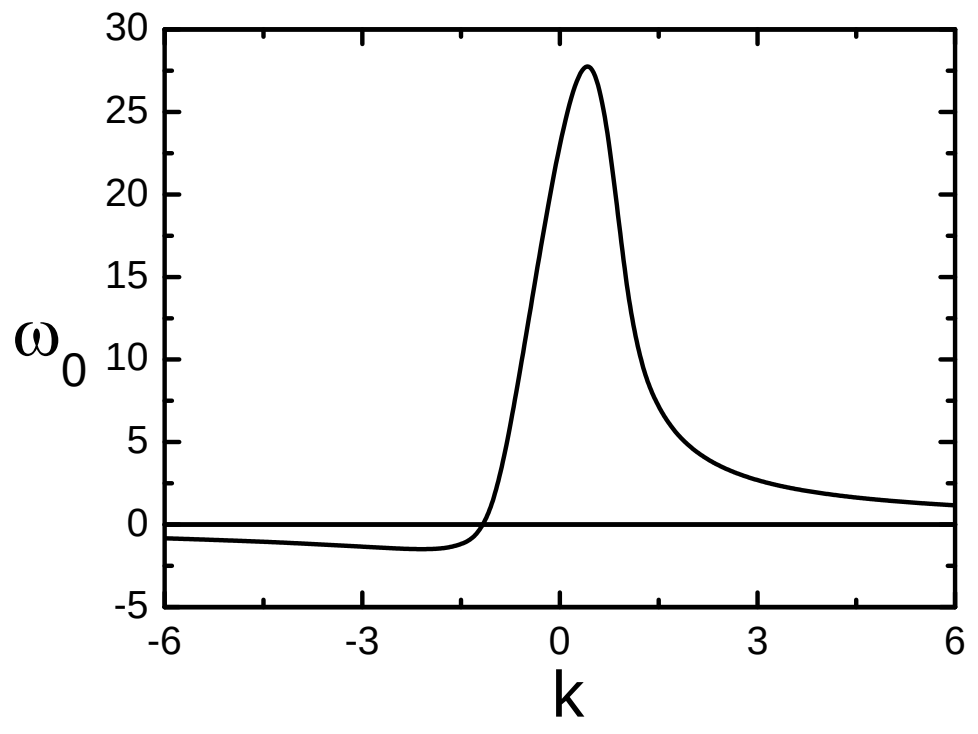

**Fig. 1** Plot of the Brans-Dicke parameter at the present time ($\omega_0$), as a function of the parameter $k$.

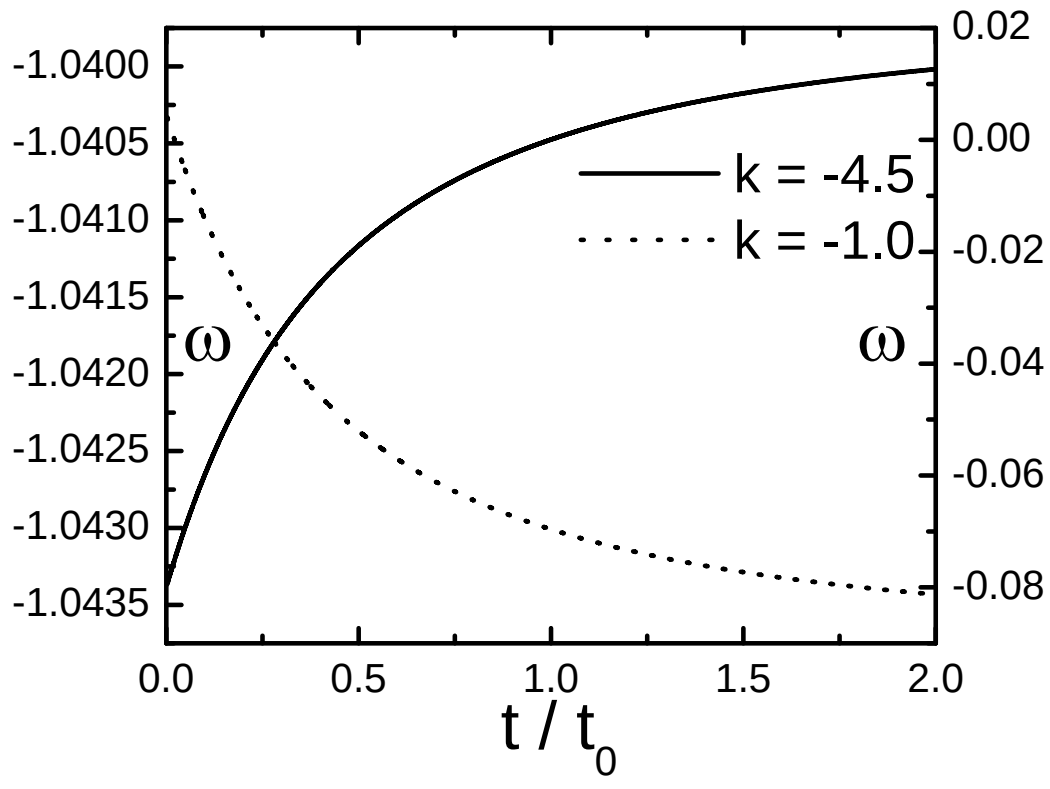

**Fig. 2** Plots of $\omega$ versus time, for two values of $k$, in its upper (*dotted*) and lower (*solid*) ranges, shown along the right and left vertical axes respectively.

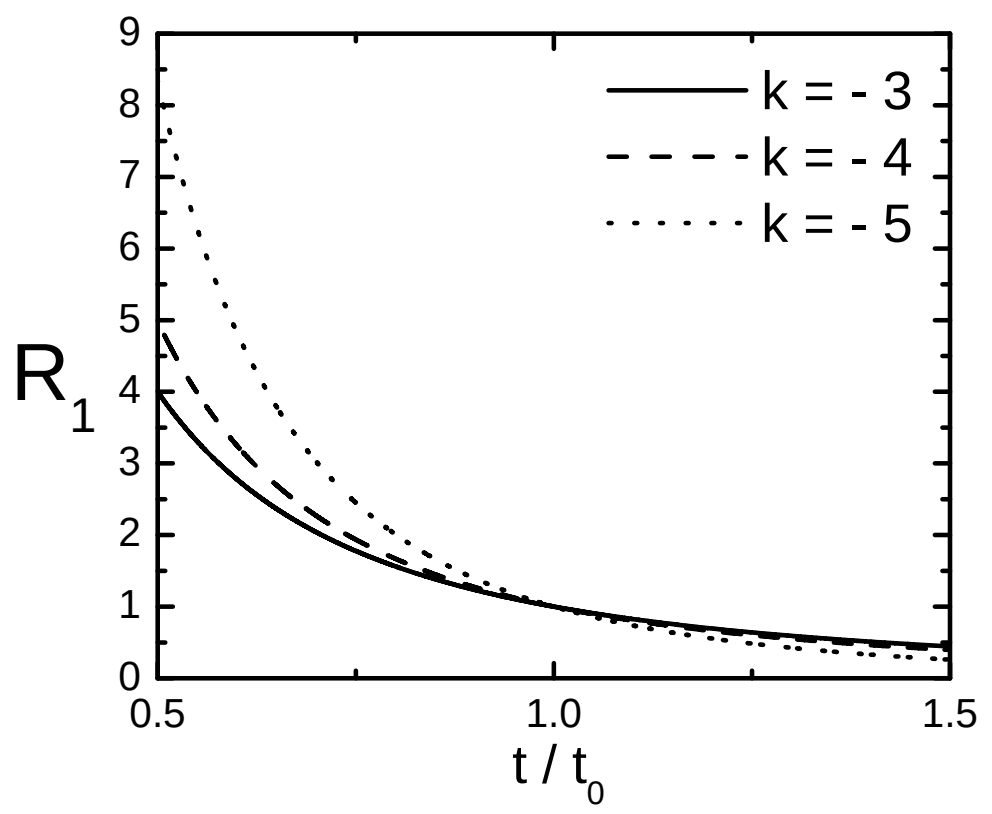

**Fig. 3** Plots of $R_1[\equiv f(t)]$ versus time for three values of $k$. Here, $R_1 = -7.20 \times 10^{-18}$ at $t = t_0$.

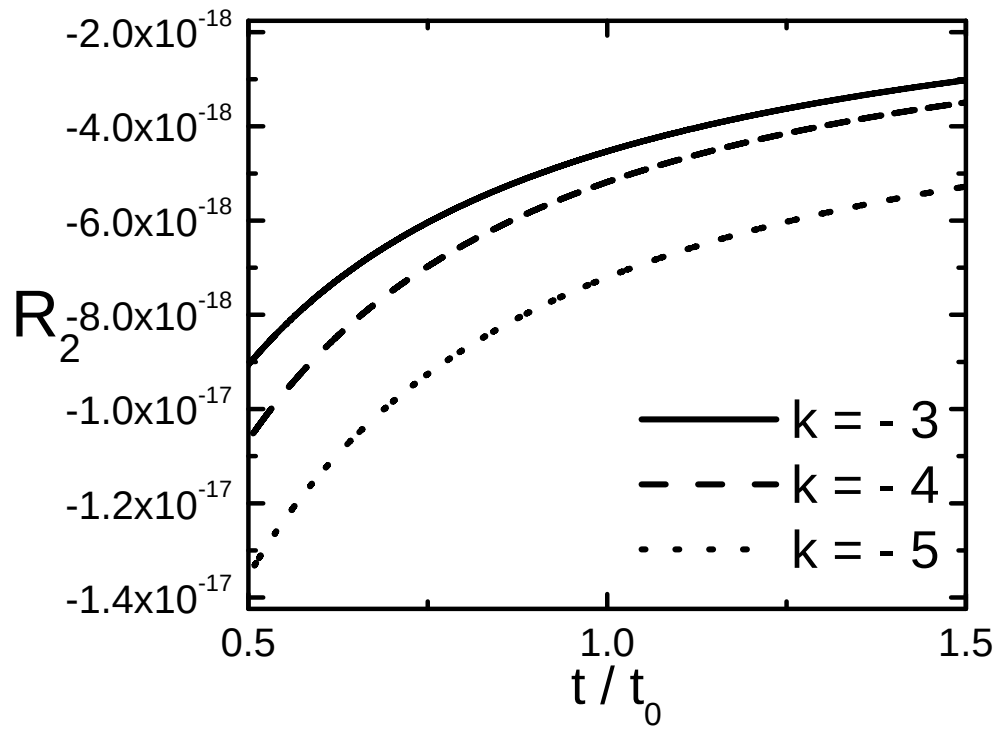

**Fig. 4** Plots of $R_2$ versus time for three values of the parameter $k$.

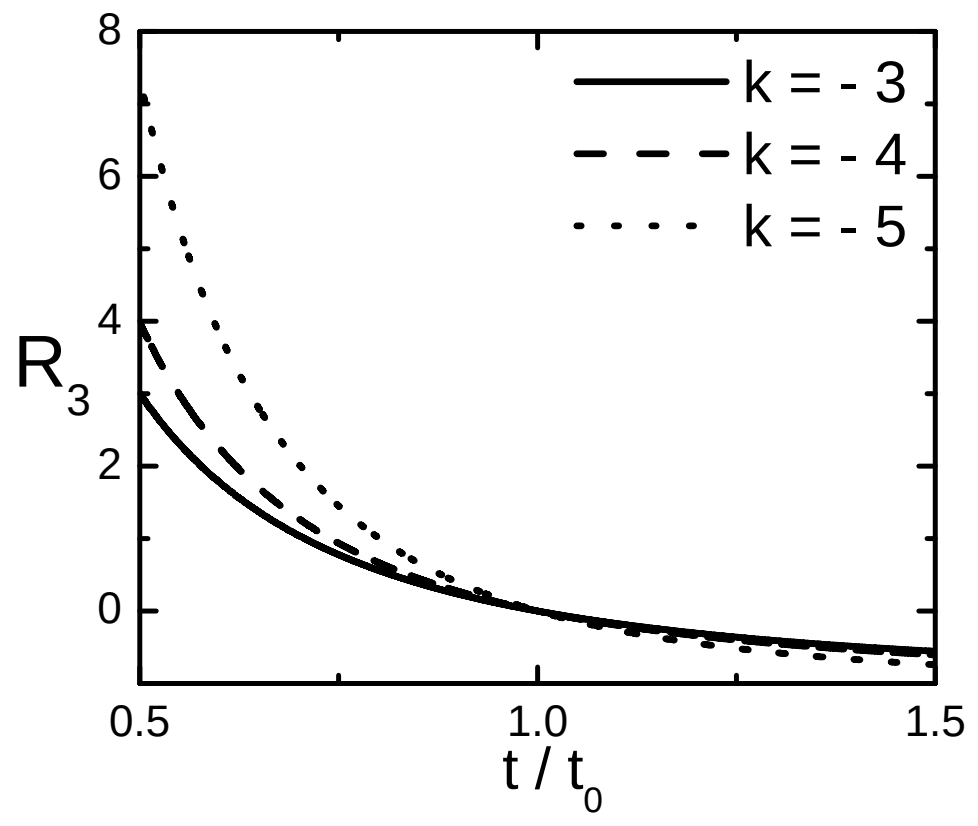

**Fig. 5** Plots of $R_3$ versus time for three values of the parameter $k$.

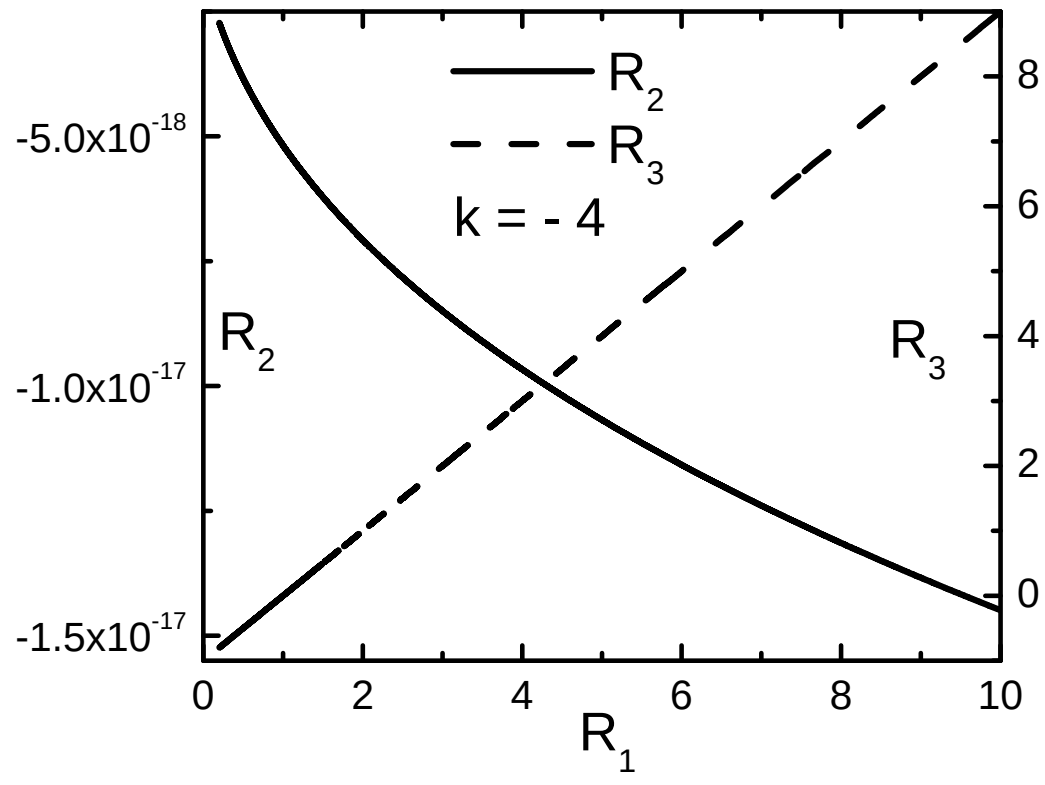

**Fig. 6** Plots of $R_2$ and $R_3$ as functions of $R_1$ [$\equiv f(t)$].

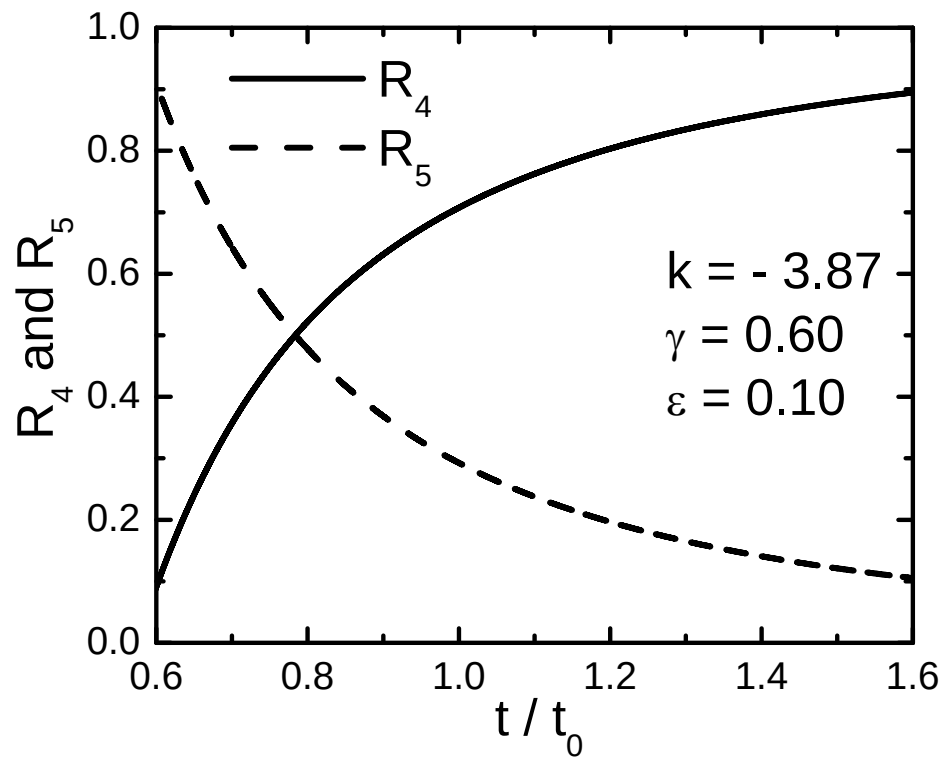

**Fig. 7** Plots of $R_4$ (*solid*) and $R_5$ (*dotted*) versus time. Conversion of matter to dark energy began at the time of $t = \gamma t_0$.

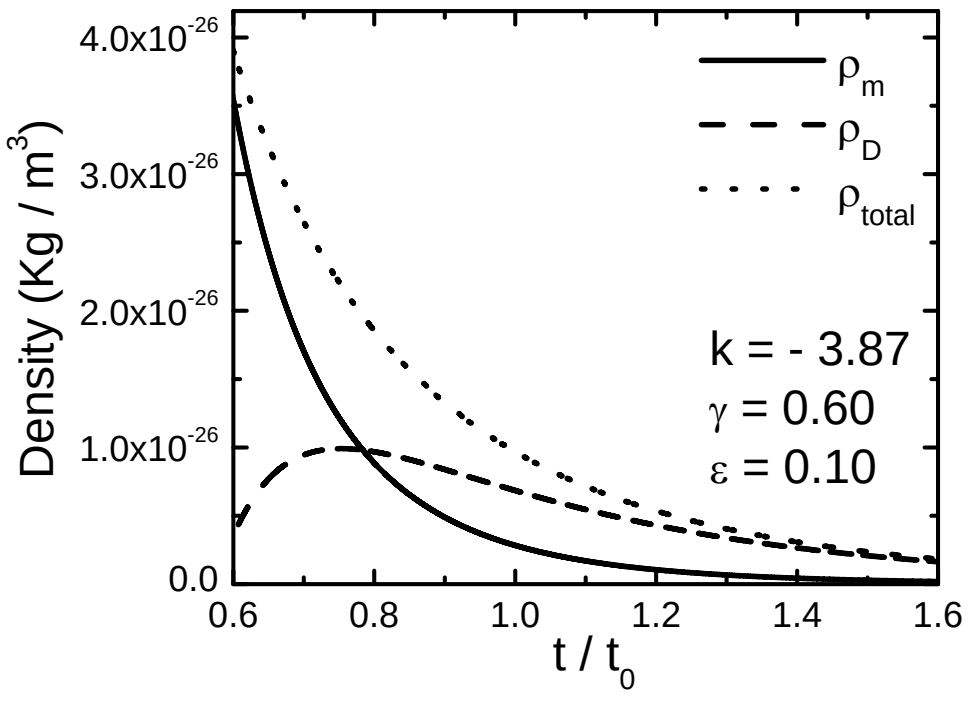

**Fig. 8** Plots of the densities of matter ($\rho_m$), dark energy ($\rho_d$) and the total matter-energy content ($\rho_{total}$) versus time.

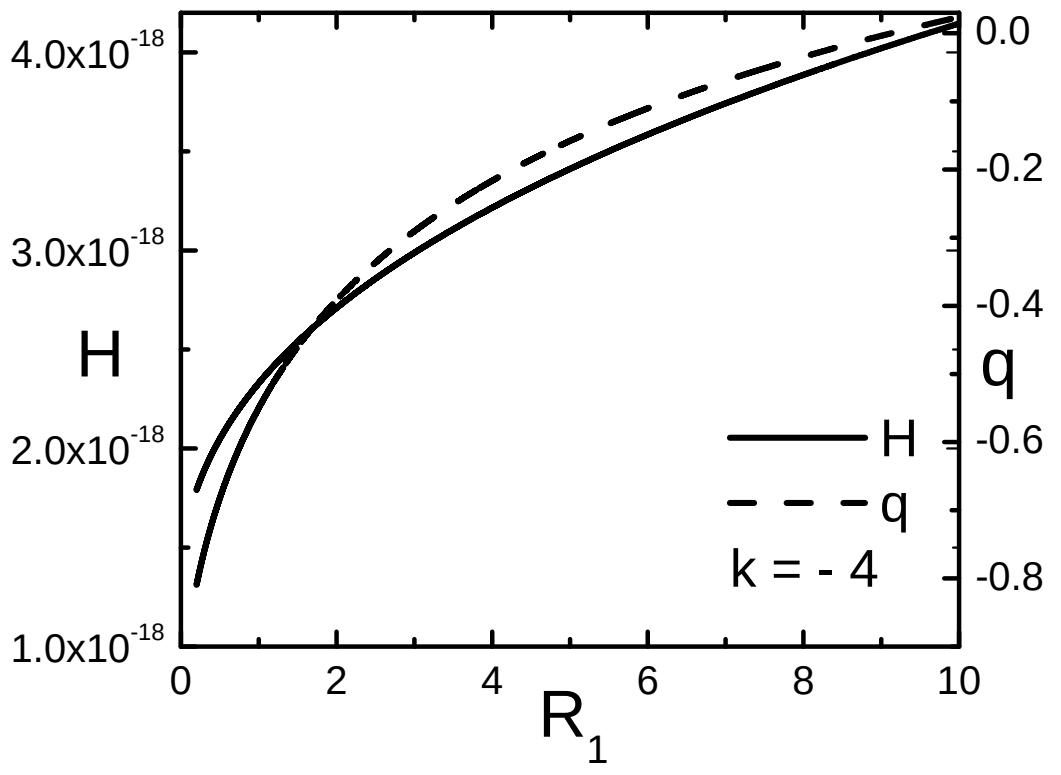

**Fig. 9** Plots of $H$ and $q$ as functions of $R_1$, along the left and right vertical axes respectively. The unit of $H$ is $sec^{-1}$.

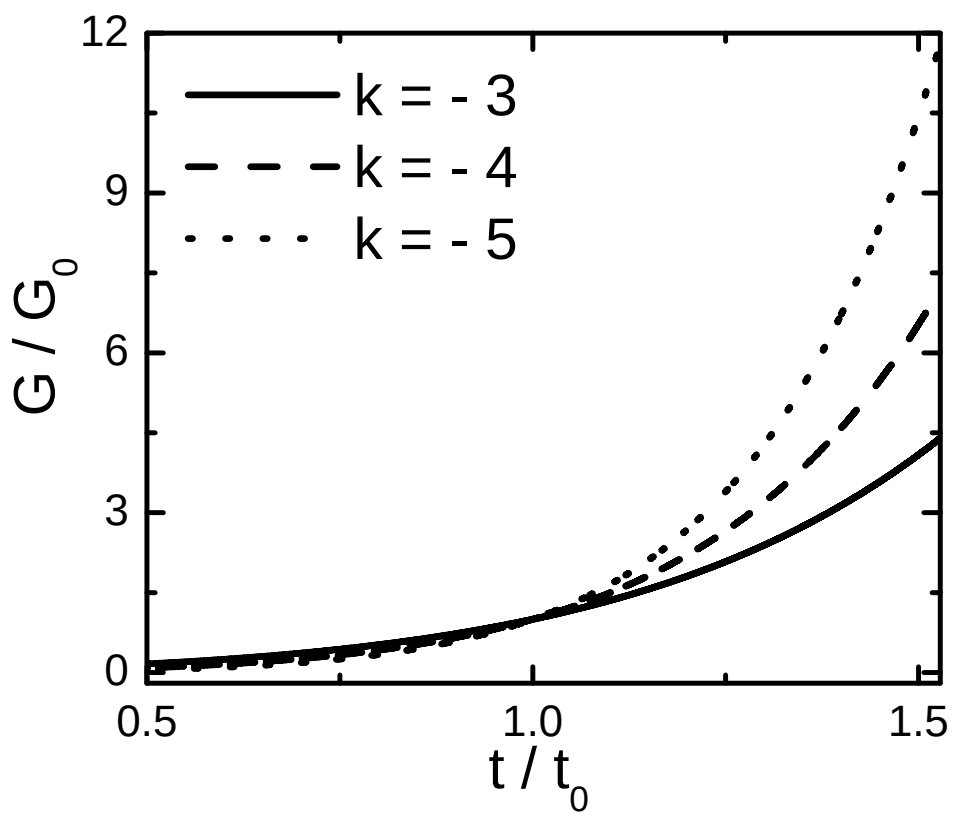

**Fig. 10** Plots of $G/G_0$ versus time for three values of the parameter $k$.

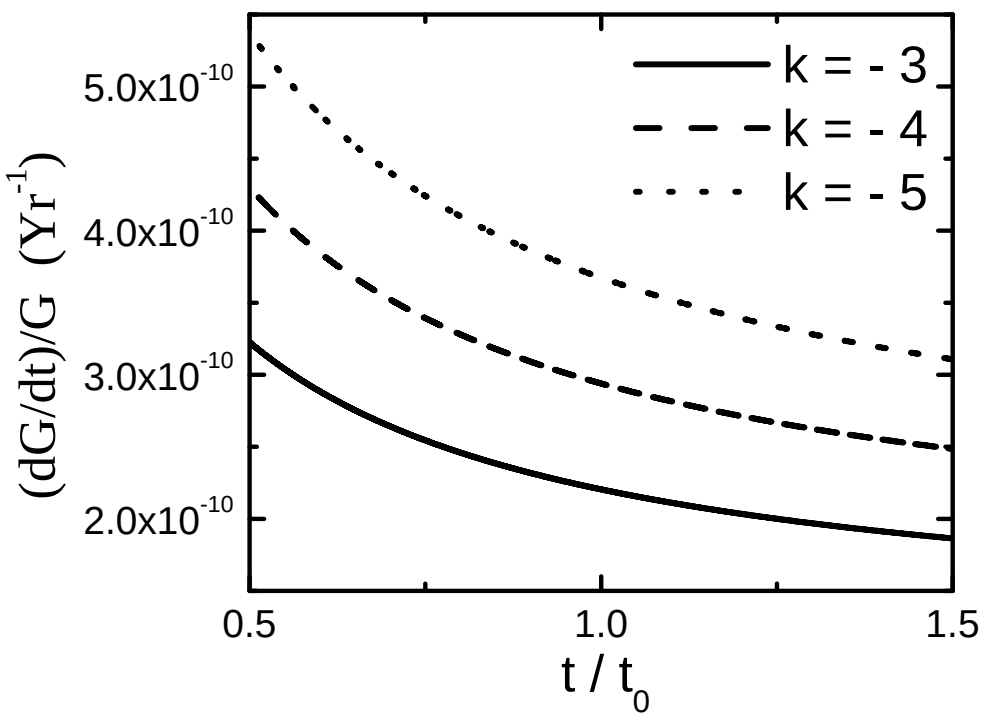

**Fig. 11** Plots of $\dot{G}/G$ versus time for three values of the parameter $k$.

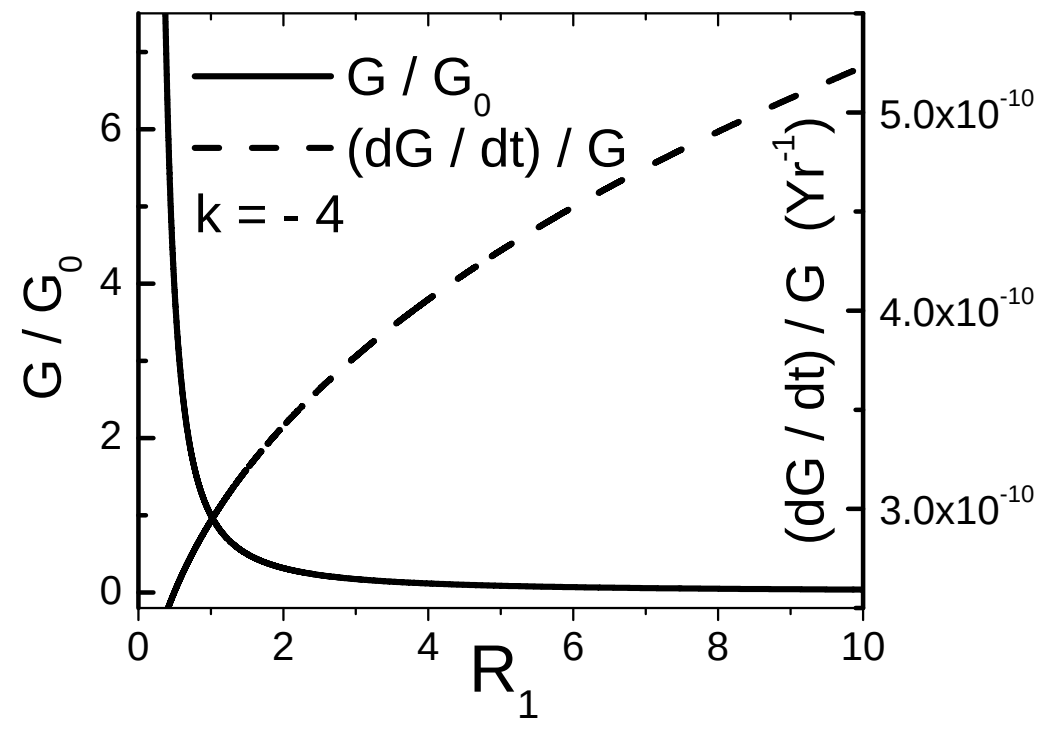

**Fig. 12** Plots of $G/G_0$ and $\dot{G}/G$ versus $R_1$, along the left and right vertical axes respectively. $\frac{\dot{G}}{G} = 7.35 \times 10^{-11}\, Yr^{-1}$ at $t = t_0$.

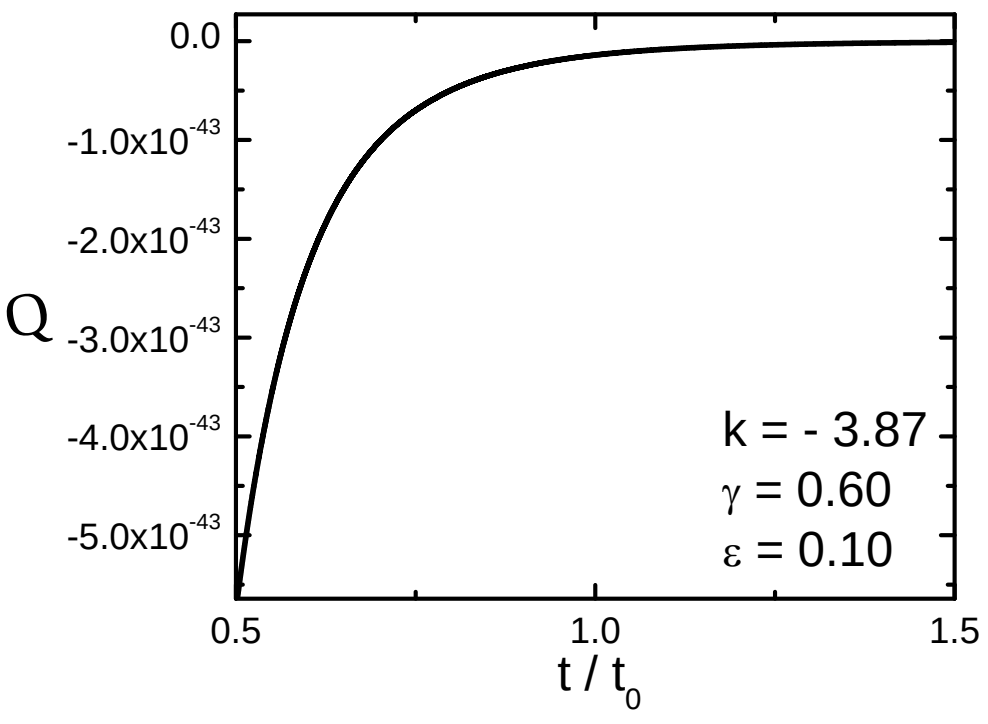

**Fig. 13** Plot of the interaction term $(Q)$ versus time. $Q = -1.4 \times 10^{-44}$ at $t = t_0$.

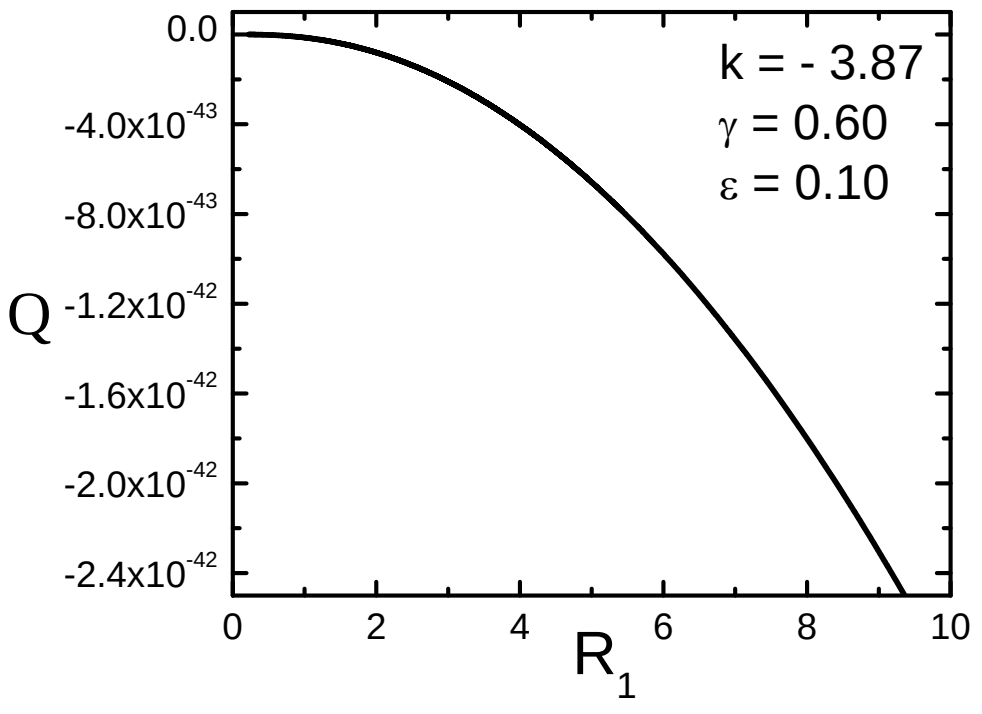

**Fig. 14** Plot of the interaction term $(Q)$ versus $R_1[\equiv f(t)]$. $Q = -1.4 \times 10^{-44}$ at $t = t_0$ when $R_1 = 1$.

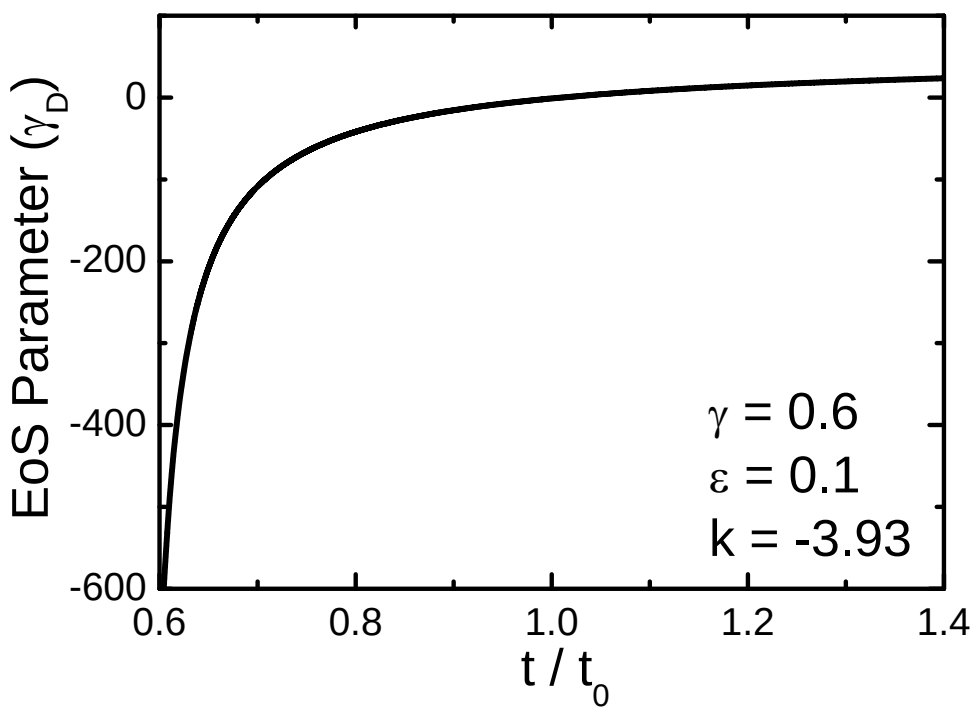

**Fig. 15** Plot of the equation of state (EoS) parameter for dark energy $(\gamma_D)$ versus time. $\gamma_D = -1.07$ at $t = t_0$.

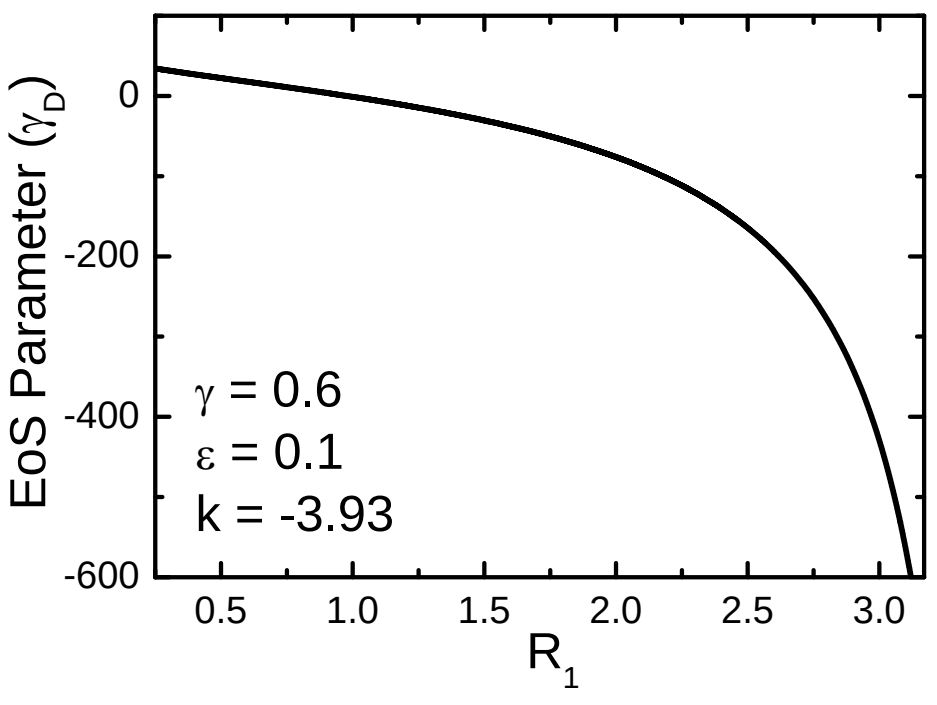

**Fig. 16** Plot of the equation of state (EoS) parameter for dark energy $(\gamma_D)$ versus $R_1[\equiv f(t)]$. $\gamma_D = -1.07$ at $t = t_0$ when $R_1 = 1$.